\documentclass[a4paper,10pt,twoside]{cpc-hepnp}

\usepackage{multicol}
\usepackage{graphicx}
\usepackage{booktabs}
\usepackage{amssymb,bm,mathrsfs,bbm,amscd}
\usepackage[tbtags]{amsmath}
\usepackage{lastpage}

\begin{document}

\fancyhead[co]{\footnotesize WANG Yu-Sa et al: Deprojection
technique for galaxy cluster considering point spread function}

\footnotetext[0]{}

\title{Deprojection technique for galaxy cluster considering point spread
function}

\author{%
      WANG Yu-Sa$^{}$\email{wangyusa@mail.ihep.ac.cn}%
\quad Jia Shu-Mei$^{}$\email{jiasm@mail.ihep.ac.cn}%
\quad Chen Yong$^{}$ } \maketitle

\address{%
(Key Laboratory for Particle Astrophysics, Institute of High Energy Physics, Chinese Academy of Sciences, Beijing 100049, China)\\
}

\begin{abstract}
We present a new method for the analysis of Abell 1835 observed by
XMM-Newton. The method is a combination of the Direct Demodulation
technique and deprojection. We eliminate the effects of the point
spread function (PSF) with the Direct Demodulation technique. We
then use a traditional deprojection technique to study the
properties of Abell 1835. Compared to that of deprojection method
only, the central electron density derived from this method
increases by 30 percent, while the temperature profile is similar.
\end{abstract}


\begin{multicols}{2}

\section{Introduction}

Being the largest and most massive celestial bodies in the universe,
galaxy clusters are the ideal laboratories for studying large-scale
structure of the Universe as well as dark matter, through which the
cosmological model\cite{lab1} could be tested. The center of galaxy
cluster is extremely complex, and the observation is susceptible to
interference. The XMM-Newton observatory is an ideal instrument for
observations of galaxy clusters for its excellent spatial and
spectral resolution. EPIC pn and MOS are the main instruments
onboard XMM-Newton designed for X-ray detection. However the PSF of
pn and MOS is not ignorable, which makes the spatial resolution
worse and causes the analysis at the central cluster inaccurate.
XSPEC could eliminate the effects of PSF through fit-ting several
spectra parameters\cite{lab2}. However, this method can not be used
with deprojection\cite{lab3} simultaneously, which is an open
problem for many years in the analysis of clusters, and the fit
models here do not consider the energy dependence of the PSF.

The direct demodulation (DD) is an effective method for image
restoring\cite{lab4}. Through Lucy iterations\cite{lab5} with
constraints of background\cite{lab6}, the restored image has better
spatial resolution, helping us learning of galaxy clusters greatly.
The DD combined with deprojection is a new method which can do PSF
correction and deprojection simultaneously without empirical model
completely. Comparing to the above method, the DD is more effective
for the sensitivity and spatial resolution, even if the statistic
and signal-to-noise ratio is bad.

Abell 1835 (z=0.2523) has a relaxed structure in which a large
cooling flow has been found by ROSAT, ASCA and
XMM-Newton\cite{lab7,lab8,lab9,lab12}. The luminosity of Abell 1835
is sufficient to DD, and it¡¯s relaxed and symmetry structure is
suitable for deprojection. In this paper we restore pn images in
different energy channel of Abell 1835 (observation ID 0098010101);
and then we employ a traditional deprojection technique applied in
several clusters successfully\cite{lab7}. Finally, the spectra
with-out PSF and projection effects are obtained. This is the first
time to deproject a galaxy cluster considering PSF.

The structure of this paper is as follows: section 2 describes PSF,
DD and processes the data of Abell 1835 with DD; section 3 presents
the deprojection method, the fit results such as the temperature
profile and the electron density, and the error estimation; then we
conclude the results in section 4.

\section{Direct Demodulation}
We can use the Lucy iteration\cite{lab5} (eq. 1) with background
constraints to restore the sub-images of Abell 1835 in 34 different
energy bands.
\begin{equation}
 f^{(r+1)}(i)=f^{(r)}(i)\sum_{k}\frac{P(k,i)d(k)}{\sum\limits_{i'}P(k,i')f^{(r)}(i')}/\sum_{k}P(k,i)
\end{equation}

where $f^{(r+1)}(i)$ can be derived from the initial
values,$f^{(0)}(i)$, with several iterations and background
constraint.

The average value of each energy band is taken for the calculation
of PSF, which is a function of energy and angle. The steps to
process a sub-image are as follows: (a) Select a region of 100x100
with the center (263, 279) as the DD region, about 500¡Á500 arcsec
square (1 bin = 5arcsec); (b) Find the positions of the gaps between
pn CCDs; (c) Select a circle, whose center is the most luminous in
the sub-image and the radius is 120 pixels; calculate rest region
without CCD gaps as back-ground; (d) Iteration with the above
background; (e) Combine 90x90 of DD region with the rest of original
sub-image into an intact sub-image, which can avoid the edge error
of DD region.

Because the DD processes the iteration with physical limit, its
error is hard to obtain through the direct error estimation. The
error depends on the PSF, background, shape of source and iteration
times. We find that the 10 times iteration is the best choice for
DD. Obvious dispersion will appear in the image when the iteration
time is more than 10, while the result is insufficiently stable when
less than 10. For estimating the errors of DD, we create 20 images
with the Poisson distribution by convolving the result $f$ and the
PSF of each energy band; DD the 20 images one by one and get 20
different values of $f$ which is $f'_{j}(j=1,...,20)$. We use the
standard deviations of $f'_{j}$ as the errors of $f$.

We take the 0.3-0.4keV sub-image as an example. The original and the
PSF-corrected images are shown in Fig.1. The counts in each ring are
calculated before and after DD. The restored image shows a much
higher counts in the central region (Fig.2).

\begin{center}
\tabcaption{ \label{tab1}  The fit results of the spectra. T1 and T2
are the temperatures in the double temperature model. A1 and A2 are
the abundances in the double temperature model. The error bars are
at the 90\% confident level..} \vspace{-3mm} \footnotesize
\begin{tabular*}{190mm}{@{\extracolsep{\fill}}ccccccccc}
\toprule ring/arcsec & T1/keV & T2/keV & A1 & A2 & Norm1/($10^{-3}cm^{-5}$) & Norm2/($10^{-3}cm^{-5}$) & $\chi^{2}_{red}$/$d_{.}o_{.}f$ & Note\\
\hline
0.0-0.25 & $1.21^{+0.19}_{-0.18}$ & $7.95^{+1.05}_{-0.89}$ & $0.21^{+0.17}_{-0.09}$ & $0.59^{+0.17}_{-0.19}$ & $0.80^{+0.32}_{-0.37}$ & $3.14^{+0.24}_{-0.25}$ & 1.340/28 & DD+deprojection \\
         & $1.11^{+0.26}_{-0.17}$ & $6.18^{+0.88}_{-0.59}$ & $0.11^{+0.07}_{-0.04}$ & $0.45^{+0.10}_{-0.10}$ & $0.60^{+0.30}_{-0.20}$ & $1.95^{+0.14}_{-0.21}$ & 1.364/28 & DD \\
0.25-0.75 & $0.62^{+0.23}_{-0.22}$ & $7.89^{+0.98}_{-0.74}$ & $0.03^{+0.04}_{-0.02}$ & $0.34^{+0.14}_{-0.12}$ & $1.21^{+0.38}_{-0.41}$ & $4.49^{+0.21}_{-0.24}$ & 1.473/28 & DD+deprojection \\
          & $0.91^{+0.20}_{-0.14}$ & $8.14^{+0.65}_{-0.52}$ & $0.04^{+0.03}_{-0.02}$ & $0.39^{+0.08}_{-0.07}$ & $1.32^{+0.30}_{-0.28}$ & $5.28^{+0.17}_{-0.24}$ & 1.630/28 & DD \\
0.75-1.5 & $1.37^{+2.09}_{-0.63}$ & $14.20^{+15.16}_{-3.12}$ & $0.03^{+0.11}_{-0.03}$ & $0.37^{+1.35}_{-0.34}$ & $1.04^{+1.04}_{-0.44}$ & $3.45^{+0.36}_{-1.31}$ & 1.031/28 & DD+deprojection \\
         & $1.09^{+0.85}_{-0.40}$ & $12.15^{+4.48}_{-1.79}$ & $0.02^{+0.04}_{-0.02}$ & $0.31^{+0.31}_{-0.20}$ & $0.89^{+0.39}_{-0.29}$ & $2.99^{+0.23}_{-0.44}$ & 0.853/28 & DD \\
         &  & $9.04^{+0.82}_{-0.69}$ &  & $0.25^{+0.15}_{-0.15}$ &  & $4.03^{+0.13}_{-0.13}$ & 1.631/31 & DD+deprojection \\
         &  & 8.27 &  & 0.23 &  & 3.45 & 2.167/31 & DD \\
1.5-2.25 &  & $7.20^{+0.94}_{-0.70}$ &  & $0.30^{+0.19}_{-0.18}$ &  & $2.65^{+0.12}_{-0.12}$ & 1.153/31 & DD+deprojection \\
         &  & $7.91^{+0.84}_{-0.80}$ &  & $0.31^{+0.17}_{-0.16}$ &  & $2.04^{+0.08}_{-0.08}$ & 1.789/31 & DD \\
\bottomrule
\end{tabular*}%
\end{center}

\begin{center}
\includegraphics[width=8cm]{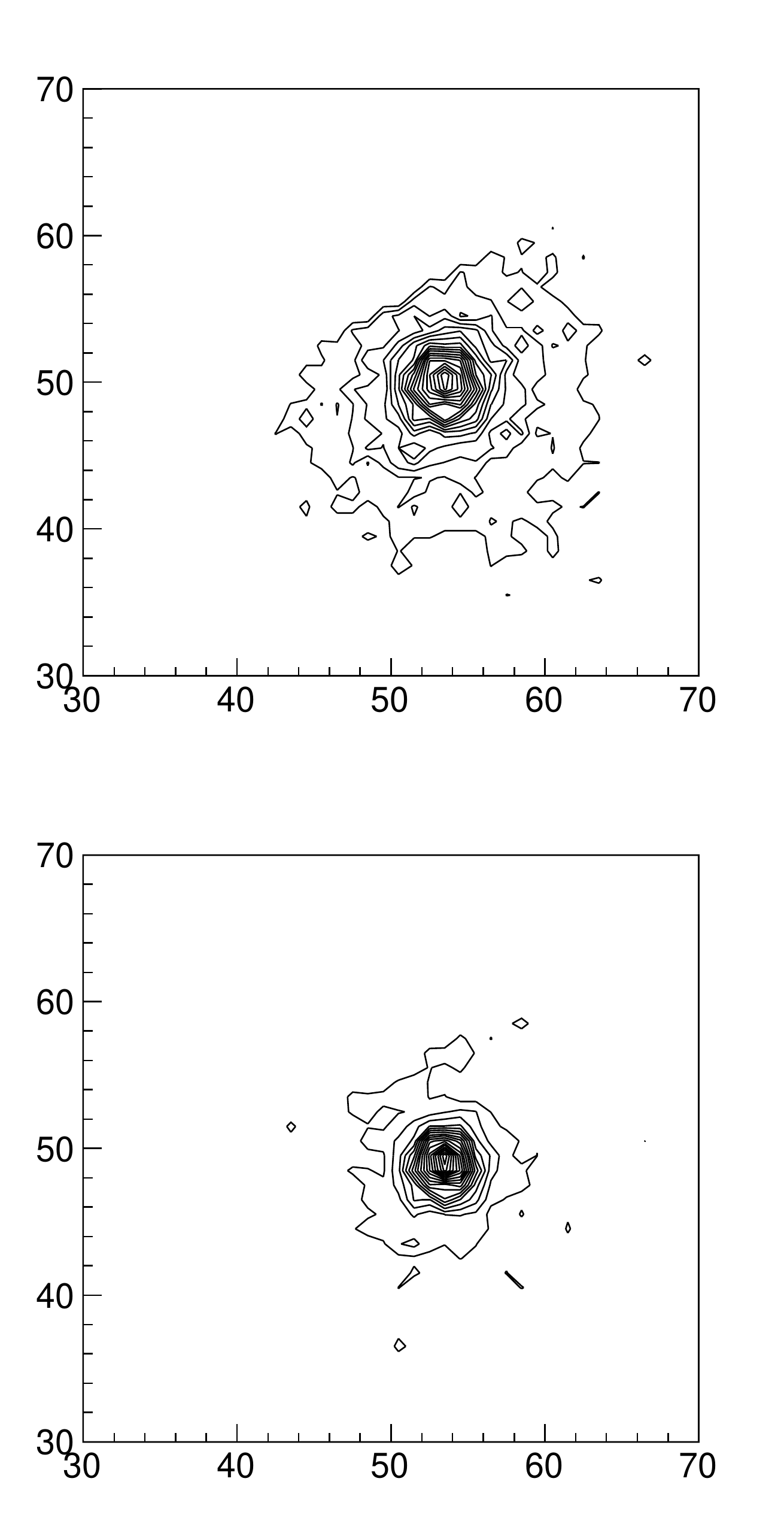}
\figcaption{\label{fig1} The top is the original image; the bottom
is the image after PSF correction. }
\end{center}

\vskip 10\baselineskip

\begin{center}
\includegraphics[width=8cm]{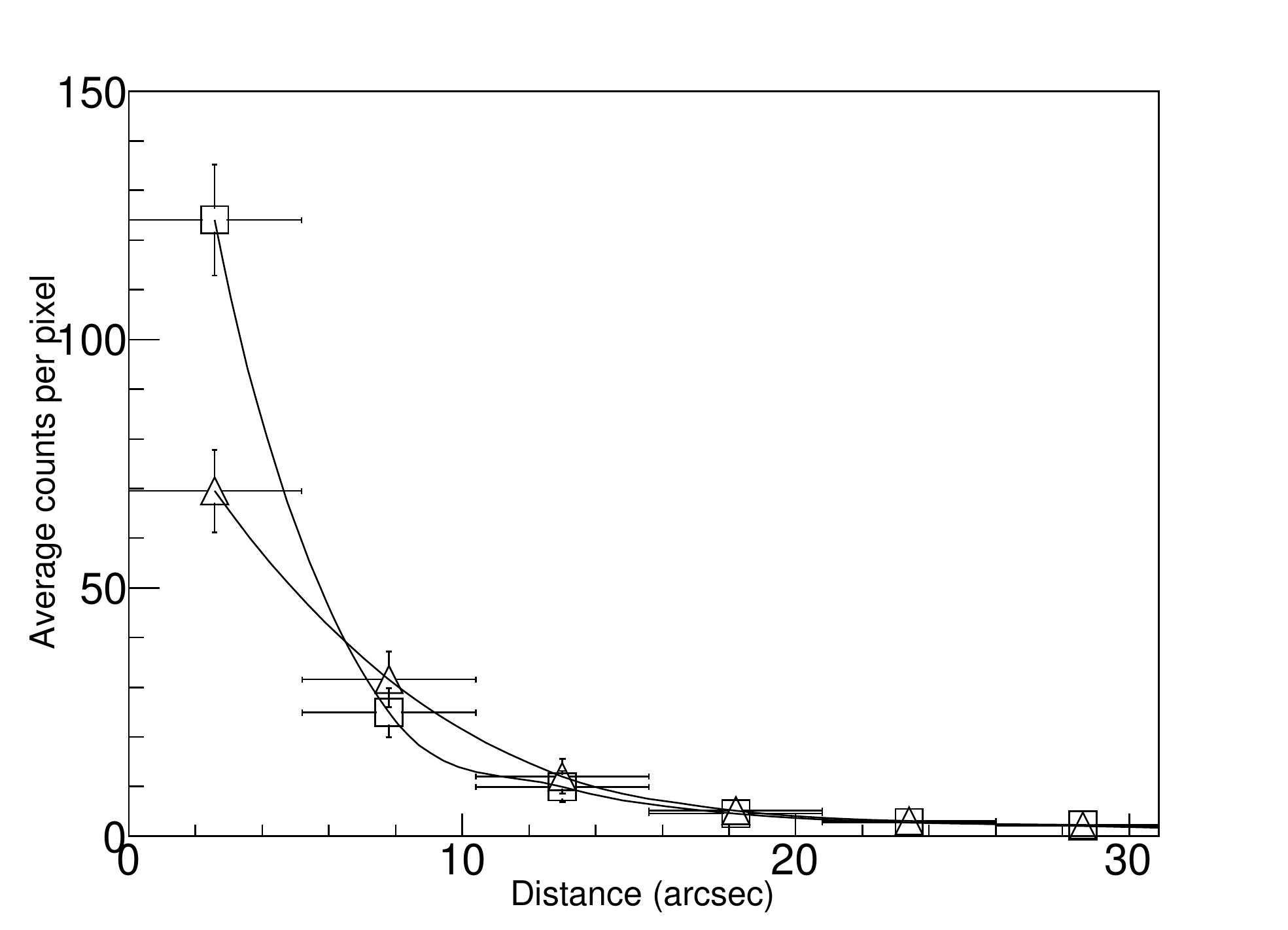}
\figcaption{\label{fig2} The average counts in each ring before and
after DD. Triangles and squares represent the original counts and
the counts after PSF correction, respectively. }
\end{center}

\section{Deprojection and spectral analysis}

\subsection{Deprojection}

Assuming the structure of Abell 1835 is spherically symmetric and
the spectra of each ring for the deprojection are the same, then the
deprojected spectra could be calculated through eliminating the
contribution from the outer rings for all spectrum components. We
divide the sub-image into 8 annular areas, whose center is located
in the most luminous pixel, and use the utmost ring $(6'-8.33')$ to
obtain the X-ray background. However, the signal in the seventh ring
is almost background, so we only consider the inner sixth rings
$(\leqslant 6')$.

\subsection{Spectra analysis} Since different version of SAS and RMF
of pn could derive different results, so we use SAS 8.0 and XSPEC
12.4.0 in this paper all the time. We analyze the DD and deprojected
spectra of pn data in XSPEC with a single and a double temperature
thermal plasma model (eq.2 and 3) as follows:
\begin{equation}
 Model_{1}=Wabs(n_{H})\times Mekal(T,z,A,norm)
\end{equation}
where $Wabs$ is a photoelectric absorption model\cite{lab10} and
$Mekal$ is a single temperature plasma emission model\cite{lab11}.
\begin{equation}
\begin{split}
 Model_{2}=Wabs(n_{H})\times (Mekal(T_{1},z,A_{1},norm_{1})\\
 + Mekal(T_{2},z,A_{2},norm_{2}))
\end{split}
\end{equation}
The latter is the double temperature model implying there are two
components in the central region with different temperatures.

Because the outer region $(3.33'-6.0')$ contains pixels out of DD
region, we ignore this ring and its adjacent ring $(2.25'-3.33')$.
We fit the three rings $(0'-1.5')$ of the central region with a
double temperature model with free abundance. It was found that the
region $(0.75'-1.5')$ can be also fitted well by a single
temperature model like the outer region $(1.5'-2.25')$.

As shown in Table1, the normalization of the high temperature
component in the innermost region is about 60 percent higher than
that of deprojection only. It means that the central electron
density increases by about 30 percent, implying a much larger
gradient of the gas density in the central region of Abell 1835.
However the resultant temperature profile is more or less the same
as that of the deprojection only, except for the innermost region
where the temperature of the high component increases significantly.

Since the profile of PSF is dependent on the photon energy (Fig.3),
the temperature of the restored spectra could be different from the
original. The PSF of high energy is steeper than that of low energy
around the several adjacent bins, especially in the inner 2 bins. In
addition, the changes of the normalization of the restored spectra
will affect the spectra of its inner rings during the deprojection
process.

\begin{center}
\includegraphics[width=8cm]{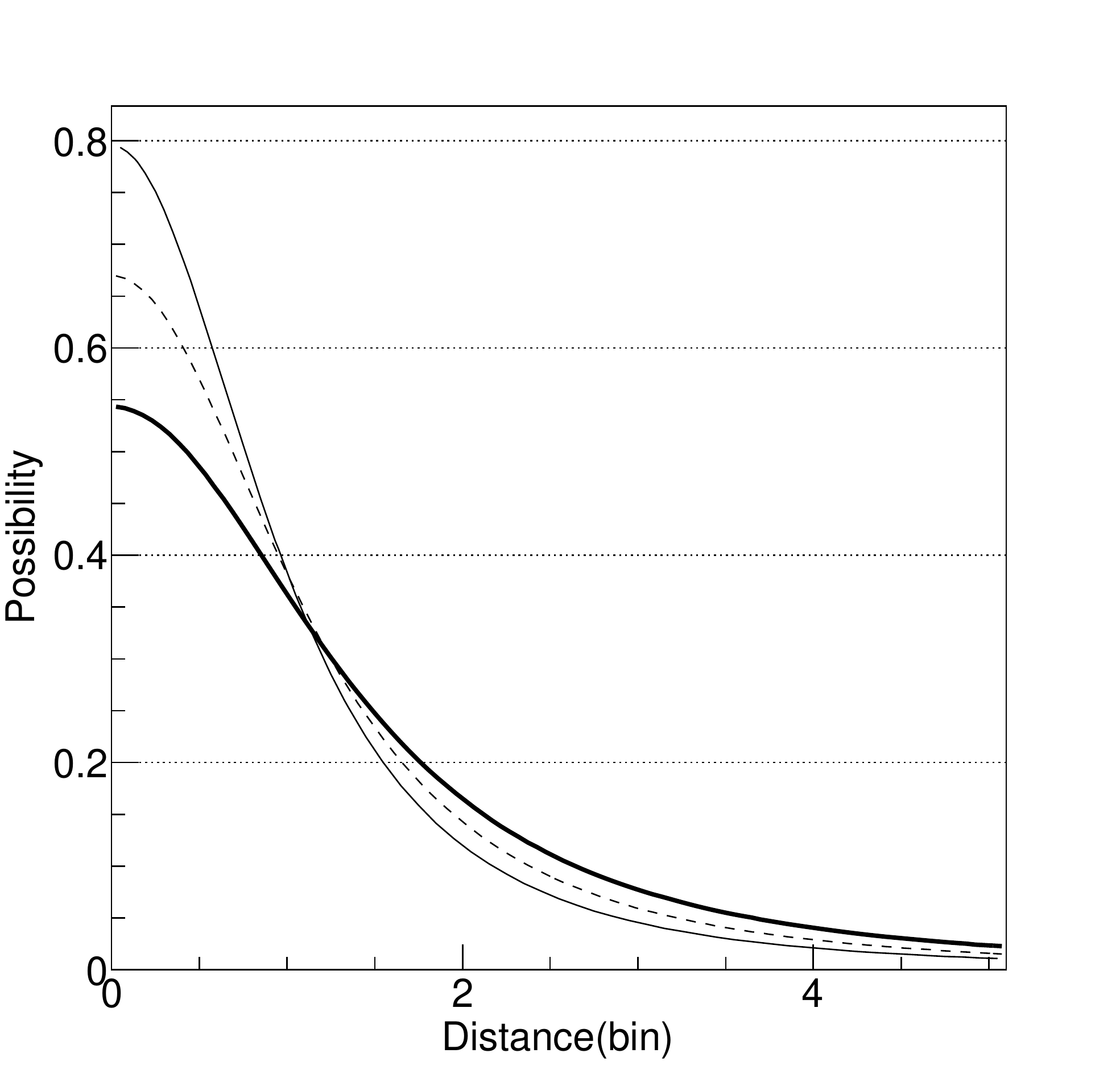}
\figcaption{\label{fig3} The PSF of pn. The heavy solid line, the
dotted line and the light solid line represent the PSF of pn of
0.3keV, 3.8keV, 7.8keV, respectively (1 bin = 5 arcsec). }
\end{center}

\begin{center}
\includegraphics[width=8cm]{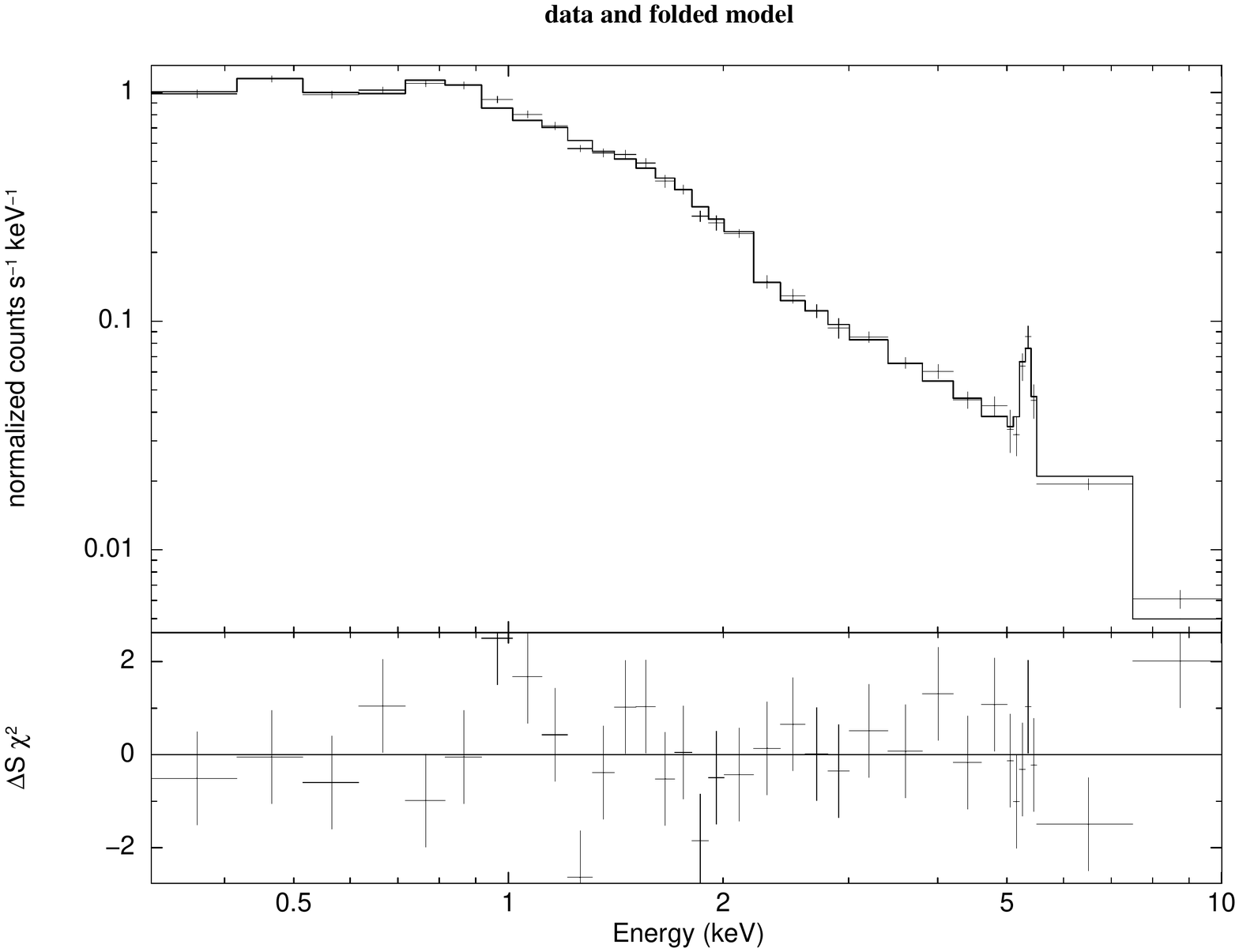}
\end{center}

\begin{center}
\includegraphics[width=8cm]{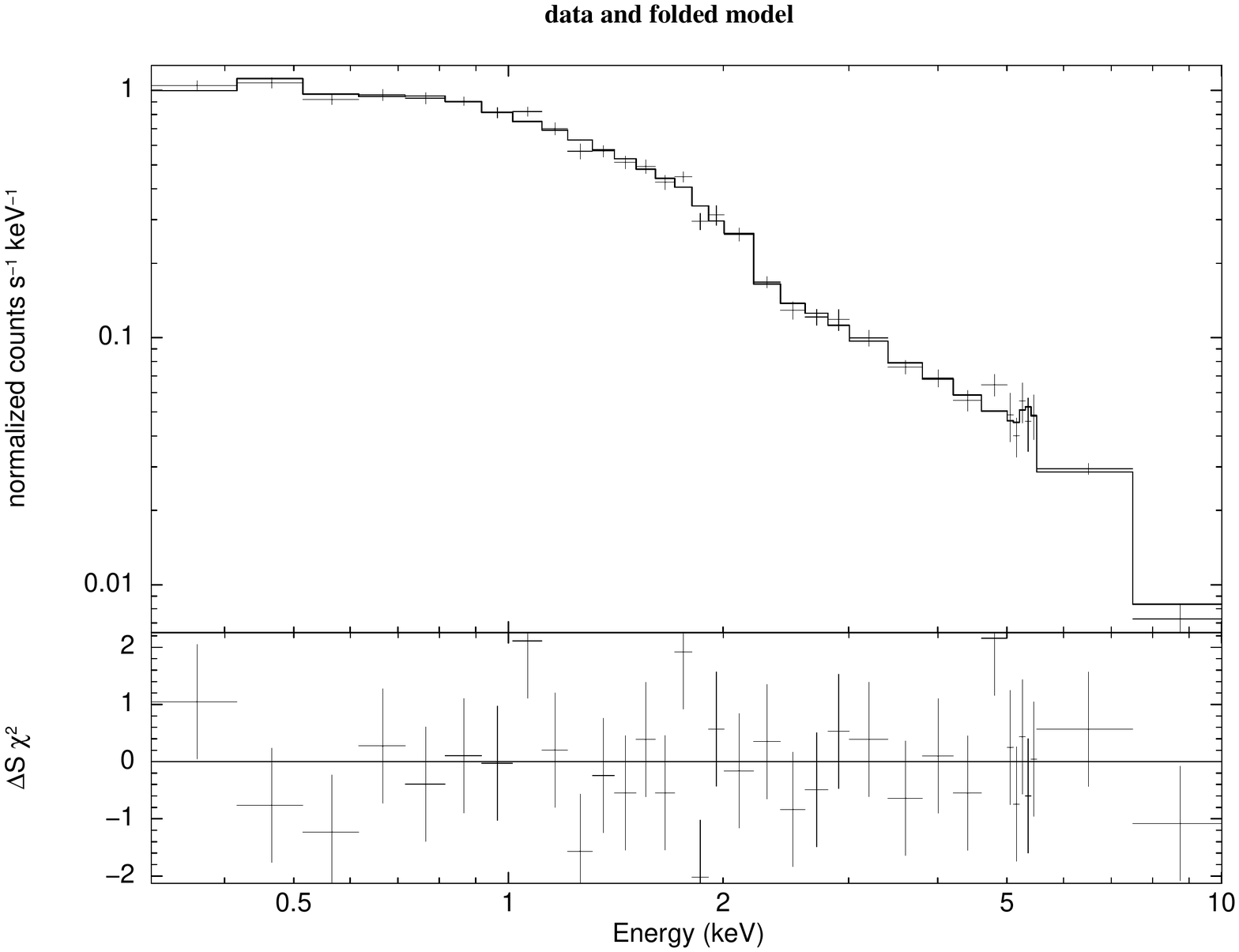}
\figcaption{\label{fig4} Spectra fitted by the double temperature
model with free abundances. The top is for the ring of $0'-0.25'$;
the bottom is for the ring of $0.25'-0.75'$. }
\end{center}

The spectra of the two innermost rings $(0'-0.25', 0.25'-0.75')$
fitted by the two temperature models are shown in Fig.4.

\section{Conclusion}
We have applied a combined DD and deprojection technique to analyze
the pn data of Abell 1835 observed by XMM-Newton observatory. The
results show that it is reasonable to analyze the cluster data with
the combined DD and deprojection technique. Comparing to the
analysis by deprojection only, we derive a similar temperature
profile but a higher central electron density. This indicates that
the effects of PSF may be not important for the temperature profile
but it should be substantial for the determination of the central
gas density as well as some other physical properties in the central
region, e.g., the cooling-flow rate.

\acknowledgments{The authors are grateful to the anonymous referees
for their insightful suggestions.  $\cdots$.}

\end{multicols}

\vspace{-2mm} \centerline{\rule{80mm}{0.1pt}} \vspace{2mm}

\begin{multicols}{2}

\end{multicols}

\vspace{5mm}

\begin{multicols}{2}

\end{multicols}
\clearpage

\end{document}